\documentclass [twocolumn, aps, showpacs] {revtex4}
\usepackage{graphicx}
\begin{document}
\draft
\title {Counterflow measurements in strongly correlated GaAs hole bilayers: evidence for electron-hole pairing}
\author {E. Tutuc, M. Shayegan}
\address{Department of Electrical Engineering, Princeton
University, Princeton, NJ 08544}
\author {D.A. Huse}
\address{Department of Physics, Princeton
University, Princeton, NJ 08544}
\date{\today}
\begin{abstract}
We study interacting GaAs bilayer hole systems, with very small
interlayer tunneling, in a counterflow geometry where equal
currents are passed in opposite directions in the two,
independently contacted layers.  At low temperatures, both the
longitudinal and Hall counterflow resistances tend to vanish in
the quantum Hall state at total bilayer filling $\nu=1$,
demonstrating the pairing of oppositely charged carriers in
opposite layers. The counterflow Hall resistance decreases much
more strongly than the longitudinal resistances as the temperature
is reduced.
\end{abstract}
\pacs{73.50.-h, 71.70.Ej, 73.43.Qt} \maketitle

In closely spaced bilayer carrier systems, the combination of
interlayer and intralayer Coulomb interaction leads to a host of
novel phenomena with no counterpart in the single-layer case
\cite{suen,eisen,murphy,lay,yang,spielman,kellogg2002,jungwirth}.
A particularly remarkable phase is the quantum Hall state (QHS)
formed at total Landau level filling factor $\nu=1$ in the limit
of zero interlayer tunneling. This state possesses unique,
interlayer phase coherence, and exhibits unusual properties, such
as Josephson-like interlayer tunneling \cite{spielman} and
quantized Hall drag \cite{kellogg2002}. The $\nu=1$ QHS can be
regarded as a condensate of excitons \cite{yang}, that is pairs of
electrons and holes in opposite layers. An alternative picture is
to view the layer degree of freedom as pseudo-spin. In this
picture, the $\nu=1$ QHS is a quantum Hall ferromagnet, where all
pseudospins point in the same direction \cite{jungwirth}.

We report magneto-transport measurements on an independently
contacted GaAs bilayer hole system in various geometries for the
current injection and voltage detection. We experimentally prove
the fundamental relations that are theoretically expected to hold
between the various transport coefficients. Focusing on the
"counterflow" configuration, we show explicitly that the
counterflow Hall resistance at $\nu=1$ tends to vanish at low
temperatures along with the longitudinal resistance. This
observation provides direct evidence that the counterflow carriers
have zero electrical charge or, equivalently, that they are
electron-hole pairs in opposite layers. We also report unexpected
behavior for the counterflow Hall resistance at $\nu=1$: as the
temperature is lowered, it drops quickly well below values of all
other resistivities, including the counterflow longitudinal
resistivity.

Our samples are Si-modulation-doped GaAs double-layer hole systems
grown on GaAs (311)A substrates. They consist of two, 150$\AA$
wide, GaAs quantum wells separated by a 75$\AA$ wide AlAs barrier.
We measured two samples from one wafer, both displaying consistent
results; here we focus on data from one sample. We used Hall bars
of $100\mu$m width, aligned along the [01$\bar{1}$] crystal
direction \cite{anisotropy}. The Hall bar mesa, shown
schematically in Fig. 1(a), has two current leads at each end, and
three leads for measuring the longitudinal and Hall voltages
across the bar. Diffused InZn Ohmic contacts are placed at the end
of each lead. We use a combination of front and back gates
\cite{eisenstein-apl} to selectively deplete one of the layers
around each contact, in order to probe the different transport
configurations of the bilayer, i.e., single layer, drag or
counterflow. As grown, the densities were $p_{T}=2.6\times10^{10}$
cm$^{-2}$ and $p_{B}=3.2\times10^{10}$ cm$^{-2}$ for the top and
bottom layers, respectively. The mobility along [01$\bar{1}$] at
these densities is approximately 20 m$^2/$Vs. Metallic top and
bottom gates were added on the active area to control the layer
densities. The measurements were performed down to a temperature
of $T=30$mK, and using low-current (0.5nA-1nA), low-frequency
lock-in techniques.

\begin{figure*}
\centering
\includegraphics[scale=0.63]{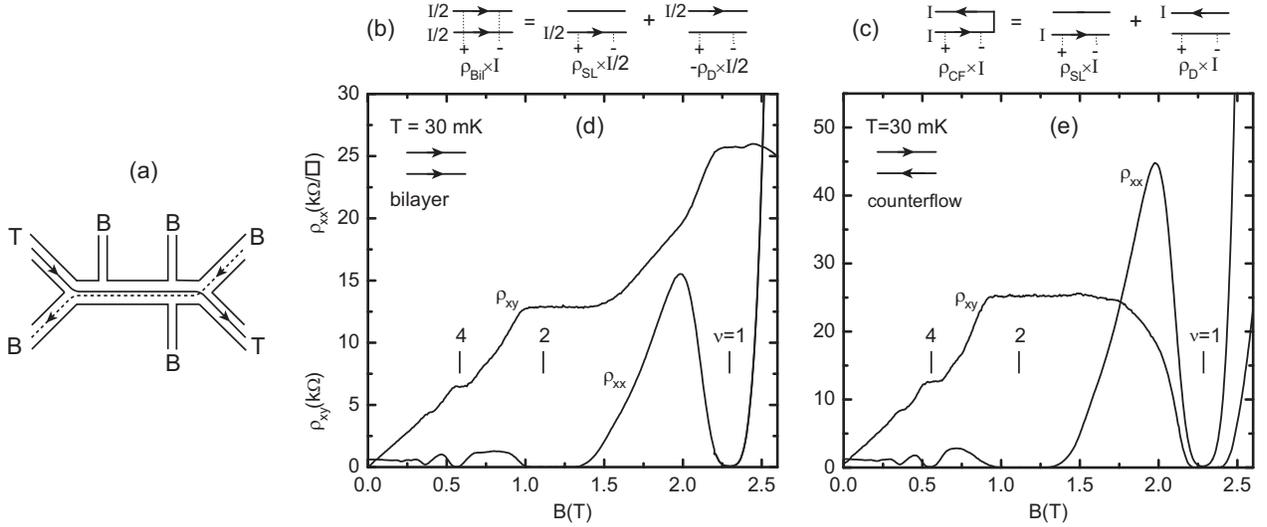}
\caption{\small{(a) Schematic representation of the Hall bar mesa
utilized in this study. Each lead contacts a particular layer
(T-top; B-bottom) in the counterflow configuration. (b,c)
Illustration of the relations between the bilayer, counterflow,
single layer and drag voltages and related transport coefficients.
(d) $\rho_{xx}$ and $\rho_{xy}$ measured at $T=30$mK in the
bilayer configuration where all contacts are made to both layers.
(e) $\rho_{xx}$ and $\rho_{xy}$ measured in the bottom layer in
the "counterflow" configuration where equal and opposite currents
are passed in the two layers. In (d) and (e) the layer densities
are $p_{B}=p_{T}=2.75\times10^{10}$ cm$^{-2}$.}}
\end{figure*}

Experimentally, several transport coefficients can be measured in
a bilayer system [Figs. 1(b,c)]. In the "bilayer" configuration
[Fig. 1(b), left], current is passed through both layers. We
define the bilayer resistivities, longitudinal and Hall, by the
corresponding voltage drops along or across the Hall bar, divided
by the total current \cite{R}. In counterflow measurements [Fig.
1(c) left] we selectively deplete one of the layers around each
contact, so that that contact is connected to only one of the
layers as indicated in Fig. 1(a). Two leads at one end of the Hall
bar are used for driving a current in and out of the sample, while
the leads at the opposite end are shorted, to ensure that the same
current, but in the opposite direction, flows in both layers.
Additionally, a current meter can be placed between the shorted
leads to measure the interlayer current leakage. The resistivities
measured in the counterflow configuration are defined as the
corresponding voltage drops along or across the Hall bar divided
by the current flowing in a single layer. This definition is
adopted so that the counterflow transport coefficients become the
same as the single layer ones, when the coupling between the
layers is negligible.

The transport coefficients, longitudinal or Hall, of a single
layer ($\rho_{SL}$) are defined by the ratio between the
corresponding voltage drops and the current flowing in that layer,
when no current flows in the opposite layer. Additionally, a
current flowing in one layer will induce a voltage drop in the
opposite layer. The drag transport coefficients ($\rho_{D}$) are
defined by the corresponding voltage drops in one (drag) layer,
divided by the current flowing in the opposite (drive) layer
\cite{drag-convention}. As depicted in Figs. 1(b,c) the bilayer
($\rho_{Bil}$) and counterflow ($\rho_{CF}$) transport
coefficients (both longitudinal and Hall) are related to the
single layer and drag coefficients:
$\rho_{Bil}=(\rho_{SL}-\rho_{D})/2$ \cite{bilayer} and
$\rho_{CF}=\rho_{SL}+\rho_{D}$. The validity of these relations is
indeed seen in our data and serves as a consistency check for our
measurements.

\begin{figure}
\centering
\includegraphics[scale=0.37]{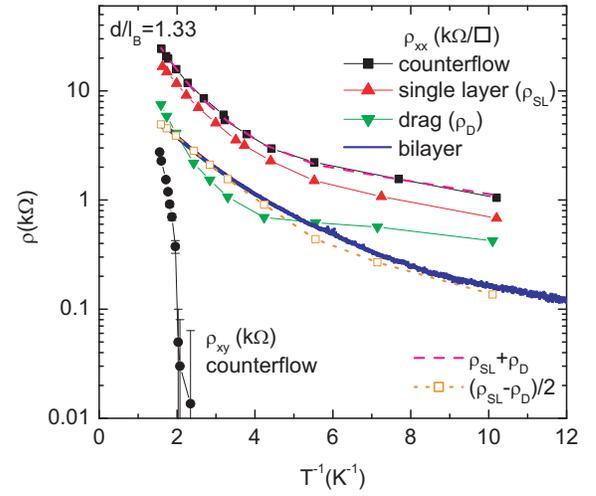}
\caption{\small{(Color online) $T$ dependence of various transport
coefficients at $\nu=1$: counterflow, single layer and drag
$\rho_{xx}$'s, and counterflow $\rho_{xy}$. Remarkably, the
counterflow $\rho_{xy}$ remains much smaller than the counterflow
$\rho_{xx}$ in the entire $T$-range. The two linear combinations
of single layer and drag transport coefficients,
$(\rho_{SL}-\rho_{D})/2$ and $\rho_{SL}+\rho_{D}$, are also
shown.}}
\end{figure}

The highlight of our results is illustrated in Figs. 1(d,e) where
the longitudinal ($\rho_{xx}$) and Hall ($\rho_{xy}$)
resistivities are shown vs perpendicular magnetic field ($B$). In
Fig. 1(d) we plot the data taken in the bilayer configuration,
when both the current and voltage leads are contacting both
layers. The data of Fig. 1(d) show a strong QHS at $\nu=1$,
evidenced by a vanishing $\rho_{xx}$ and a well developed Hall
plateau at $h/e^2\simeq25.8$k$\Omega$. A main finding of our study
is contained in Fig. 1(e) where, in the counterflow geometry,
$\rho_{xx}$ and $\rho_{xy}$ are measured on the bottom layer,
while a current of equal value flows in the {\it opposite}
direction in the other layer. The data of Fig. 1(e) show that for
the $\nu=1$ QHS, both $\rho_{xx}$ and $\rho_{xy}$ tend to vanish.
The vanishing Hall resistivity is particularly noteworthy, since
it directly proves that the counterflow current at $\nu=1$ is
transported by neutral carriers, that is, pairs of electrons and
holes in opposite layers (excitons). Indeed, an electron-hole pair
moving in one direction will create equal and opposite currents in
the two layers.

In Fig. 2 we show the $T$ dependence of $\rho_{xx}$ measured at
$\nu=1$ in different configurations: counterflow, single (bottom)
layer, drag, and bilayer. We also plot the counterflow $\rho_{xy}$
vs $T$, and show the two linear combinations of single layer and
drag longitudinal resistivities, ($\rho_{SL}-\rho_{D})/2$ and
$\rho_{SL}+\rho_{D}$. The validity of the relations
$\rho_{Bil}=(\rho_{SL}-\rho_{D})/2$ and
$\rho_{CF}=\rho_{SL}+\rho_{D}$, apparent from Fig. 2, affirms the
consistency of our data \cite{consist}. The data of Fig. 2 show a
decrease of $\rho_{xx}$ with decreasing $T$, exhibiting an
approximately exponential dependence at higher temperatures
followed by a weaker variation at lower $T$. This weaker
$T$-dependence could stem from sample disorder, a competition
between the $\nu=1$ QHS and the neighboring reentrant insulating
phase \cite{tutuc03}, or a lack of thermalization of the carriers.
The energy gap of the $\nu=1$ QHS, obtained by fitting an
exponential dependence $\rho_{xx}\propto\exp(-\Delta/2T)$ at
higher $T$ to the bilayer resistivity, is $\Delta=1.3$K, in good
agreement with previously reported results \cite{tutuc03}. A
parameter relevant to the physics of the $\nu=1$ QHS is the ratio
$d/l_{B}$ between the interlayer spacing ($d$) and the magnetic
length ($l_{B}=\sqrt{\hbar/eB}$) at $\nu=1$. This parameter, which
is a measure of the ratio of the intralayer and interlayer
interaction energies, is $d/l_{B}=1.33\pm0.1$ for the data shown
here. The error in determining $d/l_{B}$ stems from variations of
the growth rates across the sample wafer, which was not rotated
during the growth.

A striking finding of our study is the $T$-dependence of the
counterflow $\rho_{xy}$. As $T$ is increased, the counterflow
$\rho_{xy}$ remains much smaller, roughly by an order of
magnitude, than the counterflow $\rho_{xx}$. This demonstrates
that the electron-hole pairing, a prerequisite for the
stabilization of this peculiar QHS, is considerably stronger than
the $\nu=1$ QHS itself. Indeed, an exponential fit
$\rho_{xy}\propto\exp(-\Delta_H/2T)$ to the counterflow
$\rho_{xy}$ vs $T$, yields an apparent energy gap $\Delta_H=9.5K$,
much larger than the one obtained from the $T$-dependence of
$\rho_{xx}$.

To better visualize the strength of the counterflow $\rho_{xy}$ at
$\nu=1$, in Fig. 3(a) we show counterflow and bilayer $\rho_{xx}$
and counterflow $\rho_{xy}$ vs $B$, measured at $T=630$mK. The
data clearly show that the counterflow $\rho_{xx}$ is roughly an
order of magnitude larger than $\rho_{xy}$ at $\nu=1$. In Fig.
3(b), we show $\rho_{xy}$ vs $B$ traces measured for a single
(bottom) layer as well as the Hall drag. As shown in Fig. 3(a),
there is excellent agreement between the counterflow $\rho_{xy}$
and the sum of the two traces of the bottom panel. The fact that
the drag and single layer $\rho_{xy}$ are very close to the
quantized $h/e^2$ value even at this high temperature, confirms
our finding that the counterflow $\rho_{xy}$ is small.

\begin{figure}
\centering
\includegraphics[scale=0.54]{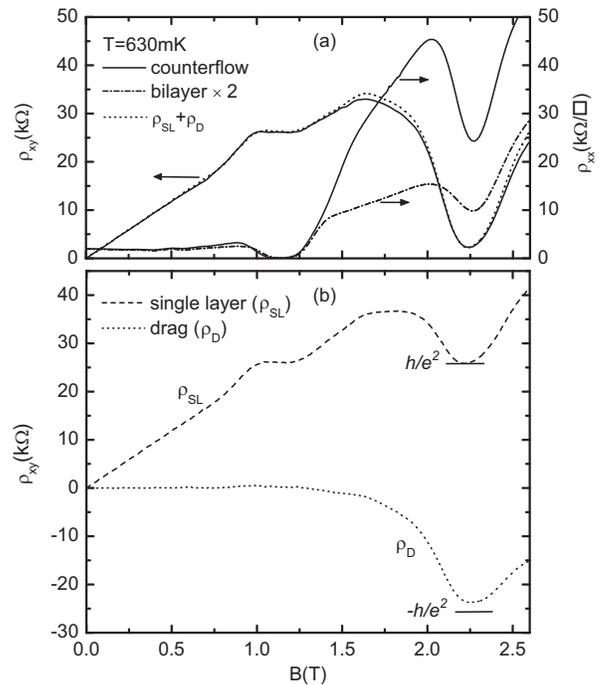}
\caption{\small{(a) Counterflow $\rho_{xx}$ and $\rho_{xy}$,
bilayer $\rho_{xx}$, and (b) single (bottom) layer $\rho_{xy}$ and
Hall drag vs $B$, all measured at $T=630$mK. As a consistency
check, the sum of these two traces is shown in panel (a) as a
dotted trace. The bilayer $\rho_{xx}$ is multiplied by 2 to
normalize it to the {\it layer} current.}}
\end{figure}

An important aspect of the drag and counterflow measurements is
the current leakage between the layers. In our counterflow
experiments, at a current of 0.5nA, the interlayer leakage at
$\nu=1$ is $\simeq4\%$ at $T=30$mK and tends to increase almost
quadratically with $T$, reaching about $8\%$ of the total current
at $T\simeq600$mK. The leakage also increases at larger currents.
Moreover, we observe smaller but comparable leakage at other
fillings, especially when the in-plane resistivity is large. Based
on these observations we believe that Josephson-like tunnelling
\cite{spielman} is not the main source of leakage at $\nu=1$. It
is unclear, however, if the leakage in our samples stems from
conventional tunnelling across the AlAs barrier between the layers
or happens at particular locations (e.g., defects in the barrier).
In the counterflow measurements, the interlayer leakage translates
to a small reduction of the actual counterflow current. In the
drag measurements, the error in the data resulting from the
interlayer leakage can be deduced by changing the ground contact
of the drag layer and recording the change in the drag signal; at
the highest temperatures ($T\simeq630$mK), where the interlayer
leakage is the largest, this error is $\pm8\%$ for $\rho_{xx}$ and
$\pm2\%$ for $\rho_{xy}$. We emphasize that these errors do not
change any of our overall conclusions.

\begin{figure}
\centering
\includegraphics[scale=0.34]{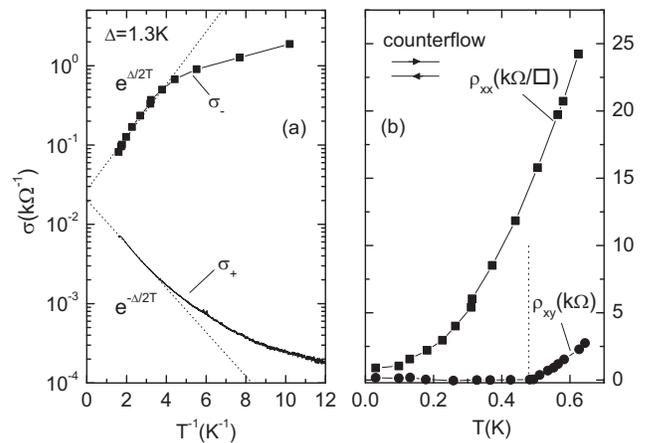}
\caption{\small{(a) Symmetric ($\sigma_{+}$) and antisymmetric
($\sigma_{-}$) conductivities vs $T^{-1}$ measured at $\nu=1$. (b)
Counterflow $\rho_{xx}$ and $\rho_{xy}$ vs $T$.}}
\end{figure}

The resistivities $\rho_{xx}$ and $\rho_{xy}$ measured in the
bilayer and counterflow configurations can be converted to
conductivities, following the usual tensor inversion. The results
represent the symmetric (bilayer), $\sigma_{+}$, and antisymmetric
(counterflow), $\sigma_{-}$, channel conductivities
\cite{factor2}. The data are shown in Fig. 4(a). $\sigma_{+}$
decreases with decreasing $T$, while $\sigma_{-}$ increases with
decreasing $T$. At higher temperatures their dependence on $1/T$
is approximately exponential with an energy gap of $\Delta=1.3K$,
followed by a saturation at lower temperatures. At the lowest
temperatures the antisymmetric conductivity is approximately four
order of magnitude larger than the symmetric one. This dependence
is a direct consequence of the fact that the counterflow
$\rho_{xy}$ is much smaller than the bilayer Hall resistivity,
which has a value of $h/e^2$ approximately, nearly independent of
temperature. We also note that, while $\sigma_{+}\ll\sigma_{-}$,
the counterflow configuration is actually more dissipative than
the bilayer one because of its higher longitudinal resistivity.

In Fig. 4(b) we show the counterflow $\rho_{xx}$ and $\rho_{xy}$
vs $T$ on a linear scale. The $T$ dependence of the counterflow
$\rho_{xy}$ is striking. As the $T$ is increased from 30mK to
0.5K, $\rho_{xy}$ remains small, even though $\rho_{xx}$ increases
considerably in this $T$ range. The counterflow $\rho_{xy}$ starts
to sharply increase at $T=0.5$K. Is this feature at about $0.5$K
the theoretically-predicted \cite{yang}, but not yet observed
Kosterlitz-Thouless (KT) transition expected for this system? If
so, the observed behavior is not as expected, since there is no
sign of a transition in the counterflow $\rho_{xx}$. The proposed
KT transition is the unbinding of vortex pairs. The vortices are
low-energy collective excitations of the correlated bilayer QHS;
they are vortices in a superfluid condensate of neutral interlayer
excitons \cite{yang}. When they are unbound, the vortices move
across the counterflow current and produce phase-slip and thus a
longitudinal resistance. We suspect that unpaired vortices remain
present in our samples to the lowest $T$ we study (perhaps due to
disorder effects) and their mobility produces the nonzero
$\rho_{xx}$. Below $T=0.5K$ their average motion is nearly
perpendicular to the counterflow current, so the counterflow Hall
angle is very near zero. In this scenario, the strong increase in
$\rho_{xy}$ above $0.5K$ is an unbinding of the neutral interlayer
excitons into uncorrelated charges that can move independently in
each layer and thus produce a counterflow Hall resistance. The
much stronger $T$ dependence of $\rho_{xy}$ than $\rho_{xx}$ would
then imply that the energy to unbind an exciton into independent
charges in each layer that produce a Hall resistance is much
larger than the energy (either an unbinding or a pinning energy)
to produce mobile vortices and thus a longitudinal resistance.

We thank R. Pillarisetty, E.A. Shaner, K. Yang, S. Girvin, and S.
Sondhi for helpful discussions, and acknowledge support by DOE and
NSF-MRSEC grants.

Note added - During the writing of this manuscript, a report on
similar experiments in a GaAs electron bilayer appeared
\cite{kellogg04}. In Ref. \cite{kellogg04} a vanishing of the
counterflow $\rho_{xx}$ and $\rho_{xy}$ at low $T$ is also
observed but, unlike our data, the $T$-dependences of these
coefficients appear to be very similar. While we do not know the
reason for this difference in behavior, we mention three factors
that distinguish our bilayer hole system from the bilayer electron
system studied in \cite{kellogg04}: (i) the estimated interlayer
tunneling in the hole system is about one order of magnitude
smaller, because of the heavier GaAs hole effective mass, (ii) the
parameter $d/l_B = 1.33$ is smaller for our sample, placing it
deeper in the $\nu=1$ bilayer quantum Hall phase (iii) the
mobility anisotropy \cite{anisotropy} could affect the current
distribution in the GaAs 2D hole samples, and possibly change the
Hall angle.


\begin{thebibliography}{10}
\bibitem{R}
The longitudinal resistivity is the measured longitudinal
resistance divided by four, the number of squares between the
voltage probes in our sample.
\bibitem{drag-convention}
We employ the following convention for the sign of the drag
resistivities: the drag resistivity, whether longitudinal or Hall,
is positive when it has the {\it opposite} sign to the voltage
drop in the drive layer.
\bibitem{bilayer}
This relation is strictly valid if the layers are identical.
\bibitem{consist}
While the our sample design does not allow us to directly measure
the top layer transport coefficients, the validity of the relation
$\rho_{Bil}=(\rho_{SL}-\rho_{D})/2$, apparent from Fig. 2, implies
that the layers have very close resitivities.
\bibitem{factor2}
The counterflow resistivities should be divided by 2, in order for
the tensor inversion to represent the antisymmetric channel
conductivities. This is because we have defined the counterflow
resistivities using the layer current and not the total current,
which is twice larger.
\end{thebibliography}
\end{document}